\documentclass[10pt,aps]{revtex4}

\usepackage{amsmath}

\begin{document}

\title{Generalized Ohm's law for relativistic plasmas}
\author{Alejandra Kandus}
\affiliation{LATO-DCET-UESC, Rodov. Ilh\'eus-Itabuna km 15 s/n,\\
CEP: 45662-000, Ilh\'eus, BA, Brazil}
\author{Christos G. Tsagas}
\affiliation{Section of Astrophysics, Astronomy and Mechanics, Department of Physics,\\
Aristotle University of Thessaloniki, Thessaloniki 54124, Greece}

\begin{abstract}
We generalise the relativistic expression of Ohm's law by studying a
multi-fluid system of charged species using the 1+3 covariant formulation of
general relativistic electrodynamics. This is done by providing a fully
relativistic, fully nonlinear propagation equation for the spatial component
of the electric 4-current. Our analysis proceeds along the lines of the
non-relativistic studies and extends previous relativistic work on cold
plasmas. Exploiting the compactness and transparency of the covariant
formalism, we provide a direct comparison with the standard Newtonian
versions of Ohm's law and identify the relativistic corrections in an
unambiguous way. The generalised expression of Ohm's law is initially given
relative to an arbitrary observer and for a multi-component relativistic
charged medium. Then, the law is written with respect to the Eckart frame
and for a hot two-fluid plasma with zero total charge. Finally, we apply our
analysis to a cold proton-electron plasma and recover the well known
magnetohydrodynamic expressions. In every step, we discuss the
approximations made and identify familiar effects, like the Biermann-battery
and the Hall effects.
\end{abstract}

\maketitle


\section{Introduction}

\label{sI} 
Relativistic plasmas play a major role in high-energy phenomena, such as
those associated with active galactic nuclei, black-hole magnetospheres,
relativistic jets, the early universe, etc. Although a complete kinetic
description of a relativistic plasma is desirable, it becomes less efficient
when the large-scale bulk plasma motions are dominant. Thus, multi-fluid
hydrodynamics is more appropriate in astrophysical systems characterized by
those kind of motions, and in this framework it is desirable to coarsen the
description to that of a one-fluid magnetohydrodynamics (MHD). The MHD
equations, which comprise of the particle conservation law, the
energy/momentum conservation laws and Maxwell's formulae, must be also
complemented by Ohm's law, namely an equation relating the induced electric
current with the electric field of the plasma. Ohm's law is directly
involved in the magnetic induction equation, used in the description of bulk
plasma dynamics. Despite its importance, however, there is still no general
consensus on the form of Ohm's law for relativistic plasmas.

Some of the previous work towards obtaining a relativistic version of the
generalised Ohm's law can be found in~\cite%
{ardavan,BF,gedalin,khanna,meier,kremer}. In general, the authors start from
a kinetic description and subsequently derive the required law in the form
of an electric-current conservation equation. The approach of~\cite{khanna}
assumes a multifluid description, but the study is then confined to a
specific black-hole metric. Here, we will derive the relativistic version of
Ohm's law for a multi-component plasma, a general metric and in the presence
of an arbitrary electromagnetic field. We will do so by means of the $1+3$
covariant approach to General Relativity and Electromagnetism (see~\cite%
{E1,EvE,BMT,TCM} for reviews and further references~\footnote{%
To a certain extend the analysis presented here resembles that given in~\cite%
{Thorne}.}). Ohm's law is given in the form of an evolution equation, along
a Fermi-propagated frame, for the spacelike component of the electric
4-current. Exploiting the mathematical compactness and physical transparency
of the covariant formalism, we arrive at an expression that allows for a
direct comparison with the existing Newtonian version of Ohm's law and
identifies the relativistic corrections in an unambiguous way. An additional
advantage is that the covariant version of Ohm's law provided here goes
beyond the cold plasma limit of the expressions obtained so far and applies
to fully relativistic fluids.

The aim of the present, first, paper is primarily to set up the mathematical
framework, discuss the physics and provide the relativistic expression of
Ohm's law in its most general possible form. The latter means that our
results can be applied to a wide variety of situations, ranging from
relativistic plasmas to astrophysics, large-scale structure and cosmology.
Thus, given any arbitrary ``background'' metric (astrophysical or
cosmological), one can easily linearise our the equations around it. In
addition, the 1+3 covariant formulation is built on irreducible kinematical
and dynamical quantities, which assigns a clear physical interpretation to
every variable in our equations. The latter should prove very advantageous
when studying nonlinear effects, particularly those outside the ideal-MHD
limit, like current-sheet formation, turbulent plasmas, magnetic-dynamo
amplification and dissipative effects. Applications of this kind will be the
subject of future work. Here, maintaining the generality of the metric and
of the electromagnetic field, we focus on a two-component fluid and apply
our multi-fluid expression of Ohm's law to a system of relativistic protons
and electrons. Adopting the Eckart frame (e.g.~see~\cite{weinberg}), allows
us to define the bulk quantities, such as the plasma velocity for example,
in a way that closely resembles those of the non-relativistic treatments.
This, in turn, helps us to discuss the common approximations in the use of
the law and to explicitly show the relativistic counterparts of very well
known Newtonian effects, like the Biermann battery and the Hall effect.

\section{Multi-component fluids}

\label{M-CFs} 

\subsection{1+3 kinematics}

\label{1+3Ks} 
Consider a pseudo-Riemannian spacetime filled with metric $g_{ab}$ of
signature ($+,+,+,-$) and introduce a timelike velocity field $u_a$
normalized so that~$u_{a}u^{a}=-1$. The tensor 
\begin{equation}
h_{ab}= g_{ab}+ u_au_b\,,  \label{h}
\end{equation}
projects orthogonal to $u_a$ and into the observers local rest-space.~%
\footnote{%
Throughout the article we use geometrised units, where $\kappa=8\pi G/c^4=1=c
$.} The $u_a$-field defines the frame of our fundamental observers and,
together with $h_{ab}$, achieves a unique 1+3 ``threading'' of the spacetime
into time and space. The same two fields are also used to define the
covariant time ans spatial derivatives of any tensor field $%
S_{ab\cdots}{}^{cd\cdots}$ according to 
\begin{equation}
\dot{S}_{ab\cdots}{}^{cd\cdots}= u^e\nabla_eS_{ab\cdots}{}^{cd\cdots} 
\hspace{10mm} \mathrm{and} \hspace{10mm} \mathrm{D}_eS_{ab\cdots}{}^{cd%
\cdots}= h_e{}^sh_a{}^fh_b{}^ph_q{}^ch_r{}^dS_{fp\cdots}{}^{qr\cdots}\,,
\label{cder}
\end{equation}
respectively (see~\cite{EvE,TCM} for reviews on the covariant formalism and
more details).

Covariantly, the kinematics of the fundamental observers are determined by
decomposing the orthogonally projected gradient of their 4-velocity field
into its irreducible parts. In particular, we have 
\begin{equation}
\mathrm{D}_bu_a= \frac{1}{3}\,\Theta h_{ab}+ \sigma_{ab}+ \omega_{ab}- \dot{u%
}_au_b\,,  \label{Dbua}
\end{equation}
where $\Theta=\nabla^au_a=\mathrm{D}^au_a$ determines the average expansion
(when positive) or contraction (when negative) of the volume element
associated with $u_a$, $\sigma_{ab}= \mathrm{D}_{\langle b}u_{a\rangle}$
indicates changes in its shape (under constant volume), $\omega_{ab}=\mathrm{%
D}_{[b}u_{a]}$ measures the rotational behaviour of $u_a$ and $\dot{u}_a=
u^b\nabla_bu_a$ is the 4-acceleration. The latter indicates the presence of
non-gravitational forces.

\subsection{Tilted 4-velocities}

\label{ssT4-Vs} 
Let us assume a mixture of fluids and define the 4-velocity $u^{a}_{(i)}$ of
the $i$-th fluid component (with~$u^{a}_{(i)}u^{(i)}_{a}=-1$) as the future
directed timelike eigenvector of the associated Ricci tensor. Then, the
tensor 
\begin{equation}
h^{ab}_{(i)}= g^{ab}+ u^{a}_{(i)}u^{b}_{(i)}  \label{hi}
\end{equation}
projects orthogonal to $u^{a}_{(i)}$. The relation between $u^{a}$ and $%
u^{a}_{(i)}$ is determined by the hyperbolic angle $\beta_{(i)}$ between the
two 4-velocity vectors. This `tilt' angle, which also determines the
peculiar motion of the $i$-th fluid component relative to $u^{a}$ (see Eq.~(%
\ref{Lboost}) below), is given by~\cite{KE} 
\begin{equation}
\cosh\beta_{(i)}= -u_{a}u^{a}_{(i)}\,,  \label{betai}
\end{equation}
with 
\begin{equation}
u^{a}_{(i)}= \gamma_{(i)}\left(u^{a}+v^{a}_{(i)}\right)  \label{Lboost}
\end{equation}
and $v_{(i)}^a$ representing the peculiar velocity of the $i$-th species
relative to the $u_{a}$-frame. Also, $\gamma_{(i)}= (1-v_{(i)}^{2})^{-1/2}$
is the Lorentz-boost factor and $v_{(i)}^2=v_{(i)}^av_a^{(i)}$. When the
tilt angle is small (i.e.~for $\beta_{(i)}\ll1$) we have $%
v_{(i)}\simeq\beta_{(i)}$. This means non-relativistic peculiar velocities
and $\gamma_{(i)}\simeq1$.

\subsection{Multi-component perfect fluids}

\label{ssM-CPFs} 
A general material medium is described by its energy momentum tensor $T^{ab}$%
, its particle flux $N^{a}$ and by the entropy flux vector $S^{a}$. The
former two quantities are conserved (i.e., $\nabla_{b}T^{ab}=0=%
\nabla_{a}N^{a}$), while the last obeys the second law of thermodynamics
(i.e., $\nabla_{a}S^{a}\geq 0$). When the strong energy condition holds, the
energy momentum tensor of a fluid accepts a unique timelike eigenvector $%
u_{T}^{a}$, normalised so that $u^{a}_{T}u^{T}_{a}=-1$~\cite{synge}. One may
also define a unitary timelike vector parallel to $N^{a}$ by $u_{N}^{a}=
N^{a}/\sqrt{-N_{a}N^{a}}$. Provided that the fluid is perfect (or in
equilibrium), $u_{T}^{a}$, $u_{N}^{a}$ and $S^{a}$ are parallel and define a
unique hydrodynamic 4-velocity vector, the rest-frame of the fluid flow (see
e.g.~\cite{israel}). This is also the only frame in which the energy
momentum tensor of the matter assumes the perfect-fluid form. When dealing
with an imperfect fluid, however, there is no longer a uniquely defined
hydrodynamic 4-velocity.

Consider a mixture of perfect fluids, where the $i$-th component has energy
density $\mu _{\left(i\right)}$, isotropic pressure $p_{\left(i\right)}$ and
moves along the timelike 4-velocity field $u_{\left(i\right)}^{a}$. Relative
to this frame, the energy momentum tensor of the individual species
decomposes as 
\begin{equation}
T_{(i)}^{ab}= \left(\mu_{(i)}+p_{(i)}\right) u_{(i)}^{a}u_{(i)}^{b}+
p_{(i)}g^{ab}= \mu_{(i)}u_{(i)}^{a}u_{(i)}^{b}+
p_{(i)}h_{\left(i\right)}^{ab}\,,  \label{Ia}
\end{equation}
with $h_{(i)}^{ab}$ given by (\ref{hi}). Also, the associated particle flux
vector is given by 
\begin{equation}
N_{(i)}^{a}=n_{(i)}u_{(i)}^{a}\,,  \label{Ib}
\end{equation}
where $n_{(i)}$ is the number density of each matter component in their own
frame.

Substituting the transformation law (\ref{Lboost}) into Eq.~(\ref{Ia})
allows us to re-express the latter with respect to the $u_{a}$-frame. The
result reads 
\begin{equation}
T_{(i)}^{ab}= \hat{\mu}_{(i)}u^{a}u^{b}+ \hat{p}_{(i)}h^{ab}+ u^{a}\hat{q}%
_{(i)}^{b}+ \hat{q}_{(i)}^{a}u^{b}+ \hat{\pi}_{(i)}^{ab}\,,  \label{Id}
\end{equation}
with $h^{ab}$ given in (\ref{h}). The above corresponds to the
energy-momentum tensor of an imperfect fluid with 
\begin{eqnarray}
\hat{\mu}_{(i)}&=& \gamma_{(i)}^{2}\left(\mu_{(i)}+p_{(i)}\right) -p_{(i)}\,,
\label{Ie} \\
\hat{p}_{(i)}&=& p_{(i)}\,,  \label{If} \\
\hat{q}_{(i)}^{a}&=&
\gamma_{(i)}^{2}\left(\mu_{(i)}+p_{(i)}\right)v_{(i)}^{a}\,,  \label{Ig} \\
\hat{\pi}_{(i)}^{ab}&=&
\gamma_{(i)}^{2}\left(\mu_{(i)}+p_{(i)}\right)v_{(i)}^{a}v_{(i)}^{b}\,.
\label{Ih}
\end{eqnarray}
We emphasise that the forms of both $\hat{p}_{(i)}$ and $\hat{\pi}_{(i)}^{ab}
$ are different from those usually found in the standard literature
(e.g.~see~\cite{KE,TCM}). In particular, $\hat{\pi}^{ab}_{(i)}$ is no longer
trace-free, which means that its does not only contain the purely
anisotropic part of the system's pressure. For the same reason $\hat{p}_{(i)}
$ is not the effective isotropic pressure of the total medium, since part of
the latter is incorporated in the $\hat{\pi}^{ab}_{(i)}$-tensor.
Neverthelees, the adopted representation has serious technical advantages,
which will allow us to go easily beyond the cold-plasma limit without
compromising the physics of our discussion.

As with the energy-momentum tensor, we may also express the particle flux of
the species relative to the $u^{a}$-frame. To be precise, combining Eqs.~(%
\ref{Lboost}) and (\ref{Ib}) we arrive at 
\begin{equation}
N_{(i)}^{a}= \hat{n}_{(i)}u^{a}+ \hat{\mathcal{N}}_{(i)}^a\,,  \label{Ii}
\end{equation}
with 
\begin{equation}
\hat{n}_{(i)}= \gamma_{(i)}n_{(i)} \hspace{15mm} \mathrm{and} \hspace{15mm} 
\hat{\mathcal{N}}_{(i)}^a= \hat{n}_{(i)}v_{(i)}^a\,,  \label{Ij}
\end{equation}
representing the associated number density and particle flux respectively.
When dealing with non-relativistic species, $\gamma_{(i)}\simeq1$ and we may
ignore terms of quadratic (and higher) order in $v_{(i)}^a$. Then, according
to (\ref{Ie})-(\ref{Ih}), we have $\hat{\mu}_{(i)}\simeq\mu_{(i)}$, $\hat{p}%
_{(i)}\simeq p_{(i)}$, $\hat{q}_{(i)}^a\simeq(\mu_{(i)}+p_{(i)})v_{(i)}^a$
and $\hat{\pi}_{(i)}^{ab}\simeq0$. Also, for $v_{(i)}^2\ll1$, relations (\ref%
{Ij}) reduce to $\hat{n}_{(i)}\simeq n_{(i)}$ and $\hat{\mathcal{N}}%
_{(i)}^a\simeq n_{(i)}v_{(i)}^a$.

\section{Multi-component charged fluids}

\label{sM-CCFs} 

\subsection{Electromagnetic fields}

\label{ssEMFs} 
The Maxwell field is covariantly characterised by the antisymmetric
electromagnetic (Faraday) tensor $F^{ab}$. Relative to an observer moving
with 4-velocity $u^{a}$, the later decomposes as (e.g.~see~\cite{E1,BMT}) 
\begin{equation}
F^{ab}= 2u^{[a}E^{b]}+ \epsilon^{abc}B_{c}\,,  \label{Ik}
\end{equation}
where $\epsilon^{abc}$ is the permutation tensor orthogonal to $u^{a}$ and $%
E^{a}=F^{ab}u_{b}$, $B^{a}=\epsilon^{abc}F_{bc}/2$ are respectively the
electric and magnetic field measured by the fiducial observer. The evolution
of these two fields is monitored by means of Maxwell's equations, which in
covariant form read~\cite{E1,BMT} 
\begin{equation}
\dot{E}^{\langle a\rangle}= -{\frac{2}{3}}\,\Theta E^a+
(\sigma^a{}_b+\omega^a{}_b)E^b+ \varepsilon^a{}_{bc}\dot{u}^bB^c+ \mathrm{%
curl}B^a- \mathcal{J}^a\,,  \label{M1}
\end{equation}
\begin{equation}
\dot{B}^{\langle a\rangle}= -{\frac{2}{3}}\,\Theta B^a+
(\sigma^a{}_b+\omega^a{}_b)B^b- \varepsilon^a{}_{bc}\dot{u}^bE^c- \mathrm{%
curl}E^a\,,  \label{M2}
\end{equation}
\begin{equation}
\mathrm{D}_aE^a= \rho- 2\omega_aB^a  \label{M3}
\end{equation}
and 
\begin{equation}
\mathrm{D}_aB^a= 2\omega_aE^a\,.  \label{M4}
\end{equation}
Note that $\mathcal{J}^{a}=h^a{}_bJ^b$ and $\rho=-u_aJ^a$ are the spatial
current and the charge density respectively, with $J_a$ representing the
4-current.\footnote{%
Angled brackets indicate the projected component of vectors (e.g.~$\dot{E}%
^{\langle a\rangle}=h^a{}_b\dot{E}^b$ -- see Eq.~(\ref{M1})) and the
orthogonally projected, symmetric and trace-free part of second-rank tensors
(e.g.~$B^{\langle a}B^{b\rangle}= B^aB^b-(B^2/3)h^{ab}$ -- see Eq.~(\ref{II0}%
)).} Also, $\omega^a= \varepsilon^a{}_{bc}\omega^{bc}/2$ is the vorticity
vector (with $\omega^{ab}=\varepsilon^{ab}{}_c\omega^c$) and $\mathrm{curl}%
v^a\equiv\varepsilon^a{}_{bc}\mathrm{D}^bv^c$ for any orthogonally projected
vector $v^a$.

The Faraday tensor obeys Maxwell's equations and also determines the
energy-momentum tensor of the electromagnetic field by means of the familiar
formula 
\begin{equation}
T_{(em)}^{ab}= F^{ac}F^{b}{}_c- \frac{1}{4}\,F^{cd}F_{cd}g^{ab}\,.
\label{Il}
\end{equation}
The latter combines with Eq.~(\ref{Ik}) to facilitate the irreducible
decomposition of $T_{\left( em\right) }^{ab}$ and a fluid description of the
Maxwell field. Thus, relative to the $u_{a}$-frame, 
\begin{equation}
T_{(em)}^{ab}= \frac{1}{2}\left(E^{2}+B^{2}\right)u^{a}u^{b}+ \frac{1}{2}%
\left(E^{2}+B^{2}\right)h^{ab}+ 2Q^{(a}u^{b)}+ \Pi^{ab}\,,  \label{II0}
\end{equation}
with $E^{2}=E^{a}E_{a}$, $B^{2}=B^{a}B_{a}$, $Q^{a}=\epsilon^{abc}E_{b}B_{c}$
and $\Pi^{ab}=-E^{\langle a} E^{b\rangle}-B^{\langle a}B^{b\rangle}$. In
other words, the electromagnetic field can be treated as an imperfect medium
with energy density $\left(E^{2}+B^{2}\right)/2$, isotropic pressure $%
\left(E^{2}+B^{2}\right)/2$, an energy flux represented by the Poynting
vector $Q^{a}$ and anisotropic stresses given by $\Pi^{ab}$~\cite{E1,BMT}.

\subsection{Conservation laws}

\label{ssCLs} 
Consider a multi-component system containing species of different nature
(e.g.~baryonic and non-baryonic matter, photons, etc) in the presence of an
electromagnetic field. The charged particles are coupled to the Maxwell
field and the mixture has a total energy-momentum tensor given by the sum $%
T_{\left( i\right) }^{ab}+T_{\left( em\right) }^{ab}$. The latter satisfies
a conservation law of the form 
\begin{equation}
\nabla _{b}T_{\left( i\right) }^{ab}-\,F^{a}{}_{b}J_{\left( i\right) }^{b}=%
\mathcal{G}_{\left( i\right) }^{a}\,,  \label{IVa}
\end{equation}%
since $\nabla _{b}T_{\left( em\right) }^{ab}=-F^{a}{}_{b}J_{\left( i\right)
}^{b}$, where $J_{\left( i\right) }^{a}$ is the electric 4-current density
of the $i$-th (charged) species. The interaction term $\mathcal{G}_{\left(
i\right) }^{a}$ represents forces other than electromagnetic. The latter, as
a consequence of the conservation of the total energy-momentum tensor, obey
the constraint 
\begin{equation}
\sum\limits_{i}\mathcal{G}_{\left( i\right) }^{a}=0\,.  \label{IVb}
\end{equation}

The timelike and spacelike parts of expression (\ref{IVa}) respectively
provide the conservation laws of the energy density and of the momentum
density of the $i$-th fluid component. Thus, by projecting (\ref{IVa}) along 
$u^{a}$ we arrive at the covariant form of the generalised continuity
equation 
\begin{equation}
\dot{\hat{\mu}}_{(i)}=-\left( \hat{\mu}_{(i)}+\hat{p}_{(i)}\right) \Theta -%
\mathrm{D}_{a}\hat{q}_{(i)}^{a}-2\dot{u}_{a}\hat{q}_{(i)}^{a}-\left( \sigma
_{ab}+\frac{1}{3}\,\Theta h_{ab}\right) \hat{\pi}_{(i)}^{ab}+\,E_{a}\mathcal{%
J}_{\left( i\right) }^{a}-\mathcal{G}_{\left( i\right) }\,,  \label{IVc}
\end{equation}%
where $\mathcal{J}_{\left( i\right) }^{a}=h^{a}{}_{b}J_{\left( i\right) }^{b}
$ and $\mathcal{G}_{\left( i\right) }=u_{a}\mathcal{G}_{\left( i\right) }^{a}
$. Note the second last term in the right-hand side of the above, which
describes alternations in the energy density of the $i$-th fluid due to the
action of the electromagnetic field. This term may be seen as representing
the familiar \emph{Joule heating} effect in a covariant manner. Recall that $%
\hat{\pi}_{(i)}^{ab}$ is not generally a trace-free tensor and therefore the
sum $h_{ab}\hat{\pi}_{(i)}^{ab}$ does not necessarily vanish. Similarly,
projecting Eq.~(\ref{IVa}) orthogonal to $u^{a}$, gives the covariant form
of the generalised Navier-Stokes equation 
\begin{eqnarray}
\left( \hat{\mu}_{(i)}+\hat{p}_{(i)}\right) \dot{u}^{a} &=&-\mathrm{D}^{a}%
\hat{p}_{(i)}-\dot{\hat{q}}_{(i)}^{\langle a\rangle }-\frac{4}{3}\,\Theta 
\hat{q}_{(i)}^{a}-\left( \sigma ^{a}{}_{b}+\omega ^{a}{}_{b}\right) \hat{q}%
_{(i)}^{b}-\mathrm{D}_{b}\hat{\pi}_{(i)}^{ab}-\hat{\pi}_{(i)}^{ab}\dot{u}_{b}
\notag \\
&&+\left( \rho _{\left( i\right) }E^{a}+\epsilon ^{a}{}_{bc}\mathcal{J}%
_{\left( i\right) }^{b}B^{c}\right) +\mathcal{G}_{\left( i\right) }^{\langle
a\rangle }\,,  \label{IVd}
\end{eqnarray}%
with $\dot{\hat{q}}_{(i)}^{\langle a\rangle }=h^{a}{}_{b}\dot{\hat{q}}%
_{(i)}^{b}$ and $\rho _{(i)}=-u_{a}J_{(i)}^{a}$ representing the charge
density of the $i$-th component.

Turning to the particle numbers and assuming that the individual number
densities are not necessarily conserved, we may write 
\begin{equation}
\nabla_{a}N_{(i)}^{a}= \mathcal{Q}_{\left(i\right)}\,,  \label{IVe}
\end{equation}
where $\mathcal{Q}_{\left(i\right)}$ indicates either an increase or a
reduction. Particle-antiparticle annihilation, for example, will reduce the
numbers, while $\sum_{i}\mathcal{Q}_{(i)}=0$ ensures overall particle
conservation. Substituting (\ref{Ii}) into the above and then using
expressions (\ref{Ij}), leads to 
\begin{equation}
\dot{\hat{n}}_{(i)}= -\left(\Theta+\mathrm{D}_{a}v_{\left(i\right)}^{a}%
\right)\hat{n}_{(i)}- v_{\left(i\right)}^{a}\mathrm{D}_{a}\hat{n}_{(i)}- 
\dot{u}_{a}v_{\left(i\right)}^{a}\hat{n}_{(i)}+ \mathcal{Q}%
_{\left(i\right)}\,,  \label{IVf}
\end{equation}
thus providing the propagation equation of the particle's number density
relative to $u_{a}$. We close this section by pointing out that both Eq.~(%
\ref{IVd}) and Eq.~(\ref{IVf}) contain implicit derivatives (temporal and
spatial) of $\gamma_{\left(i\right)}$. This is the consequence of our
decomposition choice (reflected in the set (\ref{Id})-(\ref{Ih})) and
ensures that we can treat relativistic (hot) plasmas with minimal technical
complexity.

\subsection{Electric charges and currents}

\label{ssECCs} 
The 4-current density of each charged species is related to its associated
particle-flux vector by $J_{\left(i\right)}^{a}=
eZ_{\left(i\right)}N_{\left(i\right)}^{a}$, with $e$ representing the
fundamental electric charge and $Z_{\left( i\right) }$ the atomic number of
the particles. Recalling that $N^a_{(i)}= \tilde{n}^{(i)}(u^a+v^a_{(i)})$,
relative to the $u^{a}$ frame -- see Eqs.~(\ref{Ii}), (\ref{Ij}), we have 
\begin{equation}
J_{(i)}^{a}= eZ_{(i)}\hat{n}_{(i)}\left(u^{a}+v_{(i)}^{a}\right)\,.
\label{IVg}
\end{equation}
The timelike part of the above gives the charge density of the corresponding
charged component in the fundamental frame, while its spacelike counterpart
leads to the associated 3-current. In particular, projecting (\ref{IVg})
along and orthogonal to $u_a$ we arrive at 
\begin{equation}
\rho_{\left(i\right)}= -u_{a}J_{\left(i\right)}^{a}= eZ_{\left(i\right)}\hat{%
n}_{\left(i\right)} \hspace{15mm} \mathrm{and} \hspace{15mm} \mathcal{J}%
_{\left(i\right)}^{a}= J_{\left(i\right)}^{\langle a\rangle}=
eZ_{\left(i\right)}\hat{n}_{\left(i\right)}v_{(i)}^{a}\,,  \label{IVi}
\end{equation}
respectively. Consequently, the total charge and the total 3-current are
given by the sums 
\begin{equation}
\rho= \sum_{i}\rho_{\left(i\right)}= e\sum_{i}Z_{\left(i\right)}\hat{n}%
_{\left(i\right)} \hspace{15mm} \mathrm{and} \hspace{15mm} \mathcal{J}^{a}=
\sum_{i}\mathcal{J}_{\left(i\right)}^{a}= e\sum_{i}Z_{\left(i\right)}\hat{n}%
_{\left(i\right)}v_{(i)}^{a}\,,  \label{VIa}
\end{equation}
with the former vanishing in the case of overall electrical neutrality. The
latter is a good approximation on scales larger than the Debye length of the
species, where the bulk properties of the plasma dominate. In that case $%
\sum_iZ_{(i)}\hat{n}_{(i)}=0$.

\section{Generalised Relativistic Ohm's Law}

\label{sGROL} 
Relativistic expressions of the generalised Ohm's law, in the form of a
propagation equation for the electric 3-current, appear in various versions
in the literature~\cite{ardavan,gedalin,khanna,meier}. With the exception of~%
\cite{meier}, however, all of the aforementioned studies address cold
plasmas only. Here, we use the 1+3 covariant approach to relativistic
hydrodynamics and electrodynamics to express the generalised Ohm's law in
terms of irreducible kinematical, dynamical and electrodynamical variables.
The result is an expression that closely resembles the non-relativistic
forms encountered in the standard plasma literature on one hand, while on
the other it identifies the relativistic corrections in an unambiguous way.

Covariantly, the time evolution of any given quantity is monitored by the
orthogonally projected time-derivative of the associated variable. When the
latter is spacelike, as it happens in the case of the 3-current, the
orthogonally projected time-derivative coincides with the familiar Fermi
derivative. Thus, according to (\ref{VIa}), 
\begin{equation}
\dot{\mathcal{J}}^{\langle a\rangle }=e\sum_{i}Z_{\left( i\right) }\left( 
\dot{\hat{n}}_{\left( i\right) }v_{(i)}^{a}+\hat{n}_{\left( i\right) }\dot{v}%
_{(i)}^{\langle a\rangle }\right) \,.  \label{VIb}
\end{equation}%
To obtain the full expression, one needs to replace $\dot{\hat{n}}_{\left(
i\right) }$ and $\dot{v}_{(i)}^{a}$. The evolution of $\hat{n}_{\left(
i\right) }$ is readily given by Eq.~(\ref{IVf}). The one for $v_{(i)}^{a}$,
on the other hand, is obtained by replacing (\ref{IVc}) into Eq.~(\ref{IVd}%
). Then, 
\begin{eqnarray}
\dot{v}_{(i)}^{\langle a\rangle } &=&-\frac{1}{3}\,\Theta \left(
1-v_{(i)}^{2}\right) v_{(i)}^{a}-\left( \sigma ^{a}{}_{b}+\omega
^{a}{}_{b}\right) v_{(i)}^{b}-v_{(i)}^{b}\mathrm{D}_{b}v_{(i)}^{a}+\dot{u}%
_{b}v_{(i)}^{b}v_{(i)}^{a}+\sigma _{bc}v_{(i)}^{b}v_{(i)}^{c}v_{(i)}^{a}-%
\dot{u}^{a}  \notag \\
&&-\frac{1}{\hat{M}_{(i)}}\left\{ \left[ \left( \dot{\hat{p}}_{(i)}+\,E_{a}%
\mathcal{J}_{(i)}^{a}-\mathcal{G}_{(i)}\right) v_{(i)}^{a}+\mathrm{D}^{a}%
\hat{p}_{(i)}-\left( \rho _{(i)}E^{a}+\varepsilon ^{a}{}_{bc}\mathcal{J}%
_{(i)}^{b}B^{c}\right) \right] -\mathcal{G}_{(i)}^{\langle a\rangle
}\right\} \,,  \label{dotVi}
\end{eqnarray}%
where $\hat{M}_{(i)}=\hat{\mu}_{(i)}+\hat{p}_{(i)}$. Substituting the above
result into expression (\ref{VIb}), using (\ref{IVf}) and employing some
lengthy but relatively straightforward algebra, we arrive at 
\begin{eqnarray}
\dot{\mathcal{J}}^{\langle a\rangle } &=&-\frac{4}{3}\,\Theta \mathcal{J}%
^{a}-\left( \sigma ^{a}{}_{b}+\omega ^{a}{}_{b}\right) \mathcal{J}%
^{b}+\,e\sum_{i}Z_{(i)}\left( \frac{\hat{n}_{(i)}}{\hat{M}_{(i)}}\right)
\rho _{(i)}E_{a}+\,e\sum_{i}Z_{(i)}\left( \frac{\hat{n}_{(i)}}{\hat{M}_{(i)}}%
\right) \varepsilon ^{a}{}_{bc}\mathcal{J}_{(i)}^{b}B^{c}  \notag \\
&&-e\sum_{i}Z_{(i)}\left( \frac{\hat{n}_{(i)}}{\hat{M}_{(i)}}\right) \left( 
\dot{\hat{p}}_{(i)}v_{(i)}^{a}+\mathrm{D}^{a}\hat{p}_{(i)}\right)
-e\sum_{i}Z_{(i)}\hat{n}_{(i)}\dot{u}^{a}+e\sum_{i}Z_{(i)}\left( \frac{\hat{n%
}_{(i)}}{\hat{M}_{(i)}}\right) \mathcal{G}_{(i)}^{\langle a\rangle }  \notag
\\
&&+e\sum_{i}Z_{(i)}\left[ \mathcal{Q}_{(i)}+\left( \frac{\hat{n}_{(i)}}{\hat{%
M}_{(i)}}\right) \left( \mathcal{G}_{(i)}-\,E_{b}\mathcal{J}%
_{(i)}^{b}\right) \right] v_{(i)}^{a}-e\sum_{i}Z_{(i)}\mathrm{D}_{b}\left( 
\hat{n}_{(i)}v_{(i)}^{b}v_{(i)}^{a}\right)   \notag \\
&&+e\sum_{i}Z_{(i)}\hat{n}_{(i)}\left( \frac{1}{3}\,\Theta
v_{(i)}^{2}+\sigma _{bc}v_{(i)}^{b}v_{(i)}^{c}\right) v_{(i)}^{a}\,.
\label{Ohm1}
\end{eqnarray}%
This monitors the evolution of the total 3-current, associated with a
multi-component system of non-comoving charged perfect fluids. Given that
Eq.~(\ref{Ohm1}) relates the total 3-current to the electric field, it also
provides the fully relativistic version of the generalised Ohm's law. Recall
that the \textquotedblleft hat-variables\textquotedblright\ contain the full
Lorentz-boost factors (i.e.~the $\gamma _{(i)}$'s -- see definitions (\ref%
{Ie})-(\ref{Ih}) and also (\ref{Ij}a)), which guarantees that our analysis
applies to hot as well as to cold plasmas. We also note that, in principle,
one can invert Eq.~(\ref{Ohm1}) to an expression for the electric field.
Combined with (\ref{M2}), the latter can be used to study the magnetic
component of the Maxwell field in detail.

Comparing the right-hand side of expression (\ref{Ohm1}) to that of its
Newtonian counterpart (see Eq.~(3.5.9) in~\cite{KT} for example), we can
immediately see the analogies and also locate the relativistic corrections.
Thus, the first two terms in the right-hand side of the above are due to the
relative motion of the fundamental observers. Both terms also appear in
Newtonian treatments and represent changes in the 3-current density
triggered by inertial forces. The fourth term will lead to the familiar 
\emph{Hall effect} (see \S ~\ref{ssOLHT-FPs} below), while the fifth comes
from variations in the effective `isotropic' pressure of the individual
species. Note that both the spatial and the temporal pressure gradients are
involved, with the latter treated as the relativistic correction to the
Newtonian \emph{Biermann battery effect}. The sixth term in the right-hand
side of (\ref{Ohm1}) vanishes when global electrical neutrality is imposed
and the seventh is due to particle collisions that lead to momentum transfer
between the species. The latter provide a measure of the electrical
resistivity of the total medium. The first component of the eighth term
comes from changes in the number density of the species due to
particle-creation (or annihilation) processes, while the second is triggered
by energy-density exchanges between the individual fluids and the
Joule-heating effect (see Eq.~(\ref{IVc}) in \S ~\ref{ssCLs}). The second
last term in the right-hand side of (\ref{Ohm1}) accounts for spatial
inhomogeneities in the velocities and the number densities of the individual
species, while the last is triggered by relative motion (inertial) effects.
The former has a Newtonian counterpart (e.g.~see (3.5.9) in~\cite{KT}).

\section{Hot two-fluid plasmas}

\label{sHT-FPs} 

\subsection{Eckart frame vs Landau frame}

\label{ssEFvsLF} 
A mixture of relativistic ideal fluids does not behave as an ideal medium,
which means that when considering the bulk motions of the total system we
must first specify the reference frame we are working in. Traditionally, two
choices are in order: the Eckart (or particle) frame~\cite{weinberg} and the
Landau (or energy) frame~\cite{landau}. For our multi-component fluid, the
4-velocity associated with the total particle flux is 
\begin{equation}
u_E^a= {\frac{1}{\sum_i\sqrt{-N_a^{(i)}N_{(i)}^a}}}\,\sum_iN_{(i)}^a= {\frac{%
1}{\sum_i\hat{n}_{(i)}}}\, \sum_i\left(\hat{n}_{(i)}u^a+\hat{\mathcal{N}}%
_{(i)}^a\right)\,,  \label{Eua}
\end{equation}
with the second equality resulting from decomposition (\ref{Ii}). On the
other hand, using (\ref{Id})-(\ref{Ih}), we may write the 4-velocity
associated with the total energy flux as 
\begin{equation}
u_L^a= -{\frac{1}{\sum_i\sqrt{-T_{(i)}^{ab}u_bT_{ac}^{(i)}u^c}}}\,
\sum_iT_{(i)}^{ab}u_b= {\frac{1}{\sum_i\sqrt{\hat{\mu}_{(i)}^2-\hat{q}%
_{(i)}^2}}}\, \sum_i\left(\hat{\mu}_{(i)}u^a+\hat{q}_{(i)}^a\right)\,.
\label{Lua}
\end{equation}

Observers following the Eckart frame see no particle flux, while for those
in Landau coordinates the energy flux vanishes. Thus, in our case, the
Eckart and Landau frames are defined by demanding that 
\begin{equation}
\sum_{i}\hat{\mathcal{N}}_{(i)}^{a}= 0 \hspace{15mm} \mathrm{and} \hspace{%
15mm} \sum_{i}\hat{q}_{(i)}^{a}= 0\,,  \label{Eck-Lan1}
\end{equation}
respectively. The immediate consequence is that an observer in the Eckart
frame detects a non-zero heat flux, while its Landau counterpart sees a
particle drift. On these grounds, the bulk velocities of the plasma in the
Eckart and the Landau frames are respectively defined by 
\begin{equation}
v_{E}^{a}\equiv {\frac{1}{\hat{M}}}\,\sum\limits_{i}\hat{q}_{(i)}^{a} 
\hspace{15mm} \mathrm{and} \hspace{15mm} v_{L}^{a}\equiv {\frac{1}{\hat{n}}}%
\, \sum_{i}\hat{\mathcal{N}}_{(i)}^{a}\,,  \label{Eck-Lan2}
\end{equation}
where $\hat{M}=\sum_{i}\hat{M}_{\left(i\right)}$ and $\hat{n}=\sum_{i}\hat{n}%
_{\left(i\right)}$. Both frames are physically equivalent and choosing one
against the other is a decision dictated by the particulars of the problem
in hand. In Appendix A we discuss in more detail the definition of these
frames and their non-relativist limits). Here, we shall work in the Eckart
frame because there the definitions of the bulk quantities (e.g.~the bulk
velocity of the plasma and its energy density) closely resemble their
non-relativistic associates. This will simplify our analysis and facilitate
its physical interpretation.

\subsection{Globally neutral plasmas in the Eckart frame}

\label{ssGNPEF} 
So far our analysis has been completely general, covering multi-component
systems with species of different nature (i.e.~baryons, non-baryons,
photons, etc). Most physical plasmas, however, are treated as two-fluid
mixtures of oppositely charged particles. In addition, Eqs.~(\ref{Eck-Lan2})
can now be ``inverted'' to express the velocities of the individual species
in terms of bulk variables only in two-component systems. For these reasons,
and also for mathematical simplicity, we will from now on confine to plasmas
containing one positively and one negatively charged species. All other
possible constituents (as, e.g., photons) will be considered as external.
Therefore Eq.~(\ref{IVb}) must be rewritten as 
\begin{equation}
\mathcal{G}_{+}^{a}+ \mathcal{G}_{-}^{a}= \mathcal{G}_{ext}^{a}
\label{IVb-2}
\end{equation}
where $\mathcal{G}_{ext}^{a}$ represents interactions that now are external
to our system as, for example, collisions with photons (i.e., Compton
effect), anomalous resistivity due to scattering on turbulent flows, etc
(see~\cite{gedalin} for a short account of the possible external effects).

Assuming that $v_{+}^{a}$ and $v_{-}^{a}$ are the associated velocities
relative to the fundamental $u^{a}$-frame, the corresponding energy-flux
vectors are 
\begin{equation}
q_{\pm}^{a}= \hat{M}_{\pm}v_{\pm}^{a}\,,  \label{qpm}
\end{equation}
with $\hat{M}_{\pm}=\hat{\mu}_{\pm}+\hat{p}_{\pm}= \gamma_{\pm}^{2}(\mu_{\pm
}+p_{\pm})$ (see Eqs.~(\ref{Ie})-(\ref{Ig})). Combining definition (\ref%
{Eck-Lan2}a) and expression (\ref{qpm}) we deduce that, with respect to the
Eckart frame, the bulk velocity of the system is 
\begin{equation}
v_{E}^{a}= {\frac{1}{\hat{M}}}\, \left(\hat{M}_{+}v_{+}^{a}+\hat{M}%
_{-}v_{-}^{a}\right) \,,  \label{Ebulkv1}
\end{equation}
where $\hat{M}=\sum_{\pm}\hat{M}_{\pm}$. Also, following Eq.~(\ref{VIa}b),
the total 3-current associated with our two-component system reads 
\begin{equation}
\mathcal{J}^{a}= e\left(Z_{+}\hat{n}_{+}v_{+}^{a} +Z_{-}\hat{n}%
_{-}v_{-}^{a}\right)\,,  \label{EbulkJ1}
\end{equation}
with $Z_{\pm }$ representing the atomic numbers of the positive and the
negative charges respectively.

Proceeding in line with the non-relativistic studies, we will now express
the velocities of the two charged components in terms of the bulk properties
of the plasma. The first step towards this direction is to assume overall
charge neutrality. This applies to scales larger than the Debye length of
the species, where the bulk properties of the plasma dominate. In that case, 
$Z_+\hat{n}_+= -Z_-\hat{n}_-=\hat{n}_E$ -- see definition (\ref{VIa}a) --
and expressions (\ref{IVi}), (\ref{EbulkJ1}) recast as 
\begin{equation}
\rho_{\pm}= \pm e\hat{n}_E\,, \hspace{15mm} \mathcal{J}_{\pm}^a= \pm e\hat{n}%
_Ev_{\pm}^a \hspace{15mm} \mathrm{and} \hspace{15mm} \mathcal{J}^a= e\hat{n}%
_E\left(v_+^a-v_-^a\right)\,,  \label{enrhoJ}
\end{equation}
respectively. Then, using (\ref{enrhoJ}c) we can invert Eq.~(\ref{Ebulkv1})
and arrive at 
\begin{equation}
v_{\pm}^{a}= v_E^a\pm {\frac{\hat{M}_\mp}{e\hat{n}_E\hat{M}}}\,\mathcal{J}%
^a\,,  \label{VIi}
\end{equation}
which substituted into Eq.~(\ref{enrhoJ}b) gives 
\begin{equation}
\mathcal{J}_{\pm}^a= \pm e\hat{n}_Ev_E^a+ {\frac{\hat{M}_{\mp}}{\hat{M}_++%
\hat{M}_-}}\,\mathcal{J}^a\,.  \label{EJpm}
\end{equation}
These last two results express the velocities and the 3-currents of the
individual species in terms of the corresponding bulk variables, provided
global electric neutrality holds.

Returning to the Eckart frame, we recall that $\sum_i\mathcal{N}_{(i)}^a=0$
there (see definition (\ref{Eck-Lan1})). This immediately implies that $\hat{%
n}_+v_+^a= -\hat{n}_-v_-^a$. The latter combines with relations (\ref{VIi})
and, given that the total charge density is zero, leads to 
\begin{equation}
v_E^a= {\frac{\hat{h}_+Z_++\hat{h}_-Z_-} {e\hat{n}_E\hat{h}(Z_+-Z_-)}}\,%
\mathcal{J}_a\,.  \label{Ebulkv2}
\end{equation}
Therefore, the bulk velocity of the plasma, as measured in the Eckart frame,
is colinear to total current. Finally, substituting the above into Eqs.~(\ref%
{VIi}) and (\ref{EJpm}) leads to 
\begin{equation}
v_{\pm}^a= {\frac{Z_{\pm}}{e\hat{n}_E(Z_+-Z_-)}}\,\mathcal{J}_a\,, \hspace{%
15mm} \mathrm{and} \hspace{15mm} \mathcal{J}_{\pm}^a= \pm{\frac{Z_{\pm}}{%
Z_+-Z_-}}\,\mathcal{J}_a\,,  \label{EvJpm}
\end{equation}
respectively.

\subsection{Ohm's law for hot two-fluid plasmas}

\label{ssOLHT-FPs} 
The last two relations monitor the kinematics of the species in terms of the
bulk quantities and with respect to the Eckart (particle) frame, provided
overall electrical neutrality holds. That aside, no other restriction has
been imposed and the individual charged species are completely general and
fully relativistic. In what follows we will use the results of \S ~\ref%
{ssGNPEF} to obtain Ohm's law for a relativistic two-fluid plasma. Thus,
recalling that zero total charge implies $Z_{+}\hat{n}_{+}=-Z_{-}\hat{n}_{-}$
and employing some straightforward algebra, Eq.~(\ref{Ohm1}) recasts to 
\begin{eqnarray}
\dot{\mathcal{J}}^{\langle a\rangle}&=& -\frac{4}{3}\,\Theta\mathcal{J}^{a}-
\left(\sigma^{a}{}_{b}+\omega^{a}{}_{b}\right)\mathcal{J}^{b}+ {\frac{e^{2}%
\hat{n}_{E}^{2}(\hat{M}_{+}+\hat{M}_{-})} {\hat{M}_{+}\hat{M}_{-}%
}}\,E^{a}+ {\frac{e^{2}\hat{n}_{E}^{2}(\hat{M}_{+}+\hat{M}_{-})} 
{\hat{M}_{+}\hat{M}_{-}}}\, \varepsilon^{a}{}_{bc}v_{E}^{b}B^{c} 
\notag \\
&&-{\frac{e\hat{n}^{E}(\hat{M}_{+}-\hat{M}_{-})} {\hat{M}_{+}%
\hat{M}_{-}}}\, \varepsilon^{a}{}_{bc}\mathcal{J}^{b}B^{c}- {\frac{e\hat{n}%
_{E}}{\hat{M}_{+}\hat{M}_{-}}} \left[\hat{M}_{-}\left(\dot{\hat{p}}%
_{+}v_{+}^{a} +\mathrm{D}^{a}\hat{p}_{+}\right) -\hat{M}_{+}\left(\dot{\hat{p%
}}_{-}v_{-}^{a} +\mathrm{D}^{a}\hat{p}_{-}\right)\right]  \notag \\
&&+\frac{e\hat{n}_{E}}{\hat{M}_{+}}\, \mathcal{G}_{ext}^{\langle a\rangle}- {%
\frac{e\hat{n}_{E}(\hat{M}_{+}+\hat{M}_{-})} {\hat{M}_{+}\hat{M}_{-}}}\,%
\mathcal{G}_{-}^{\langle a\rangle}+ \frac{e\hat{n}_{E}}{\hat{M}_{+}}\,%
\mathcal{G}_{ext}v_{+}^{a}- {\frac{e\hat{n}_{E}}{\hat{M}_{+}\hat{M}_{-}}}\,%
\mathcal{G}_{-} \left(\hat{M}_{-}v_{+}^{a}+\hat{M}_{+}v_{-}^{a}\right) 
\notag \\
&&-e\mathcal{Q}_{-}\left(Z_{+}v_{+}^{a}-Z_{-}v_{-}^{a}\right)- {\frac{e^{2}%
\hat{n}_{E}^{2}} {\hat{M}_{+}\hat{M}_{-}}}\, E_{b}\left(\hat{M}%
_{-}v_{+}^{b}v_{+}^{a} +\hat{M}_{+}v_{-}^{b}v_{-}^{a}\right)- e\mathrm{D}_{b}%
\left[\hat{n}_{E} \left(v_{+}^{a}v_{+}^{b}-v_{-}^{a}v_{-}^{b}\right)\right] 
\notag \\
&&+e\hat{n}_{E}\left[\left({\frac{1}{3}}\,\Theta
v_{+}^{2}+\sigma_{bc}v_{+}^{b}v_{+}^{c}\right)v_{+}^{a} -\left({\frac{1}{3}}%
\,\Theta v_{-}^{2} +\sigma_{bc}v_{-}^{b}v_{-}^{c}\right)v_{-}^{a}\right]\,.
\label{rOhm1}
\end{eqnarray}
For the physical interpretation of all the right-hand side terms and a
comparison with their Newtonian counterparts, we refer the reader to the
discussion given in \S ~\ref{sGROL} after Eq.~(\ref{Ohm1}). Relative to that
expression, we have kept the velocities of the individual species and
replaced the associated 3-currents by means of (\ref{enrhoJ}b). The only
exception was when dealing with the fourth term in the right-hand of (\ref%
{Ohm1}), where the 3-currents of the species were replaced using relation (%
\ref{EJpm}). The latter has allowed us to include the magnetic convection
term and the Hall term in the right-hand side of Eq.~(\ref{rOhm1})
explicitly. Keeping the fluid velocities, on the other hand, will help
obtain the non-relativistic limit of the above (see \S ~\ref{sOLCP-EPs}
below). Also note that the condition of global charge neutrality has removed
the 4-acceleration term from (\ref{Ohm1}). Finally, for the interactions we
wrote (\ref{IVb-2}) as $\mathcal{G}_{+}^{\langle a\rangle}+\mathcal{G}%
_{-}^{\langle a\rangle}= \mathcal{G}_{ext}^{\langle a\rangle}$, $\mathcal{G}%
_{+}+\mathcal{G}_{-}=\mathcal{G}_{ext}$ and assumed that $\mathcal{Q}_{+}+%
\mathcal{Q}_{-}=0$. The first two conditions respectively guarantee momentum
and energy-density conservation in absence of external interactions, while
the last ensures that the overall particle number is preserved.

Expression (\ref{rOhm1}) can be given in a variety of forms depending on the
problem in hand. For example, using relations (\ref{EvJpm}a) and (\ref{EvJpm}%
b) we can recast the right-hand side of (\ref{Ohm1}) in terms of the total
3-current. Alternatively, one can employ Eq.~(\ref{Ebulkv2}) to express
everything in terms of the bulk velocity. Here we will do the former, while
maintaining the condition of global charge neutrality. To compactify the
results, we also introduce the auxiliary bulk variables 
\begin{equation}
\hat{M}= \hat{M}_{+}+ \hat{M}_{-}\hspace{15mm} \mathrm{and} \hspace{15mm} 
\hat{\Delta}= \hat{M}_{+}- \hat{M}_{-}\,,  \label{hMhDel}
\end{equation}
which immediately imply that $4\hat{M}_{+}\hat{M}_{-}= \hat{M}^{2}-\hat{%
\Delta}^{2}$. Employing these relations, Eqs.~(\ref{Ohm1}), (\ref{rOhm1})
transform into 
\begin{eqnarray}
\dot{\mathcal{J}}^{\langle a\rangle}&=& -\frac{4}{3}\,\Theta\mathcal{J}^{a}-
\left(\sigma^{a}{}_{b}+\omega^{a}{}_{b}\right)\mathcal{J}^{b}+ {\frac{4e^{2}%
\hat{n}_{E}^{2}\hat{M}} {(\hat{M}^{2}-\hat{\Delta}^{2})}}%
\,E^{a}+ {\frac{2e\hat{n}_{E}[\hat{M}(Z_{+}+Z_{-}) -\hat{\Delta}%
(Z_{+}-Z_{-})]} {(\hat{M}^{2}-\hat{\Delta}^{2})(Z_{+}-Z_{-})}}\,
\varepsilon^{a}{}_{bc}\mathcal{J}^{b}B^{c}  \notag \\
&&-{\frac{2[\hat{M}(\dot{\hat{p}}_{+}Z_{+}-\dot{\hat{p}}_{-}Z_{-}) -\hat{%
\Delta}(\dot{\hat{p}}_{+}Z_{+}+\dot{\hat{p}}_{-}Z_{-})]} {(\hat{M}^{2}-\hat{%
\Delta}^{2})(Z_{+}-Z_{-})}}\,\mathcal{J}^{a}- {\frac{2e\hat{n}_{E}}{\hat{M}%
^{2}-\hat{\Delta}^{2}}}\, \left[\hat{M}\left(\mathrm{D}^{a}\hat{p}_{+} -%
\mathrm{D}^{a}\hat{p}_{-}\right) -\hat{\Delta}\left(\mathrm{D}^{a}\hat{p}%
_{+} +\mathrm{D}^{a}\hat{p}_{-}\right)\right]  \notag \\
&&+\frac{2e\hat{n}_{E}}{\hat{M}+\hat{\Delta}}\, \mathcal{G}_{ext}^{\langle
a\rangle}- {\frac{4e\hat{n}_{E}\hat{M}}{\hat{M}^{2}-\hat{\Delta}^{2}}}\, 
\mathcal{G}_{-}^{\langle a\rangle}+ {\frac{2Z_{+}\mathcal{G}_{ext}} {%
\left(Z_{+}-Z_{-}\right)\left(\hat{M}+\hat{\Delta}\right)}}\, \mathcal{J}%
_{a}- {\frac{2\left[\hat{M}\left({Z_{+}}+{Z_{-}}\right) -\hat{\Delta}\left({%
Z_{+}}\,-{Z_{-}}\right)\right]\mathcal{G}_{-}} {\left(Z_{+}-Z_{-}\right)
\left(\hat{M}^{2}-\hat{\Delta}^{2}\right)}}\,\mathcal{J}_{a}  \notag \\
&&-{\frac{(Z_{+}+Z_{-})\mathcal{Q}_{-}}{n_{E}}}\,\mathcal{J}^{a}+ {\frac{2[%
\hat{\Delta}(Z_{+}^{2}-Z_{-}^{2}) -\hat{M}(Z_{+}^{2}+Z_{-}^{2})]} {%
(\hat{M}^{2}-\hat{\Delta}^{2})(Z_{+}-Z_{-})^{2}}}\, E^{b}%
\mathcal{J}_{b}\mathcal{J}^{a}- {\frac{Z_{+}+Z_{-}}{e(Z_{+}-Z_{-})}}\,%
\mathrm{D}^{b} \left({\frac{1}{\hat{n}_{E}}}\,\mathcal{J}_{b}\mathcal{J}%
^{a}\right)  \notag \\
&&+{\frac{Z_{+}^{3}-Z_{-}^{3}} {e^{2}\hat{n}_{E}^{2}(Z_{+}-Z_{-})^{2}}}
\left({\frac{1}{3}}\,\Theta\mathcal{J}^{2} +\sigma^{bc}\mathcal{J}_{b}%
\mathcal{J}_{c}\right)\mathcal{J}^{a}\,,  \label{rOhm2}
\end{eqnarray}
with $\hat{M}^{2}-\hat{\Delta}^{2},\,Z_{+}-Z_{-}\neq 0$. Relations (\ref%
{rOhm1}) and (\ref{rOhm2}) provide the relativistic (1+3 covariant) version
of the generalised Ohm's law, with respect to the Eckart frame, when applied
to a mixture of two hot and interacting charged fluids. We also remind the
reader that these results have been obtained under the assumption of global
electrical neutrality.

\subsection{Relativistic particle-antiparticle plasmas}

\label{ssRP-APs} 
Expression (\ref{rOhm2}) is particularly useful when dealing with
particle-antiparticle pairs. An electron-positron plasma, for a example, has 
$Z_{\pm}=\pm1$ and $M_{+}=M_{-}$. The latter means that $\hat{M}=2\hat{M}%
_{\pm}$, $\hat{\Delta}=0$ and $\hat{M}^{2}-\hat{\Delta}^{2}=\hat{M}^{2}=4%
\hat{M}_{\pm}^{2}$ (see definitions (\ref{hMhDel})). In such an environment
Eq.~(\ref{rOhm2}) simplifies to 
\begin{eqnarray}
\dot{\mathcal{J}}^{\langle a\rangle}&=& -\frac{4}{3}\,\Theta\mathcal{J}^{a}-
\left(\sigma^{a}{}_{b}+\omega^{a}{}_{b}\right)\mathcal{J}^{b}+ {\frac{4e^{2}%
\hat{n}_{E}^{2}}{\hat{M}}}\,E^{a}- {\frac{2\dot{\hat{p}}}{\hat{M}%
}}\,\mathcal{J}^{a}- {\frac{4e\hat{n}_{E}}{\hat{M}}}\,\mathcal{G}%
_{-}^{\langle a\rangle}  \notag \\
&&-{\frac{1}{\hat{M}}}\, E^{b}\mathcal{J}_{b}\mathcal{J}^{a}+ {%
\frac{1}{2e^{2}\hat{n}_{E}^{2}}} \left({\frac{1}{3}}\,\Theta\mathcal{J}^{2}
+\sigma^{bc}\mathcal{J}_{b}\mathcal{J}_{c}\right)\mathcal{J}^{a}\,,
\label{rp-aOhm}
\end{eqnarray}
and contains no Hall effect (since $\hat{\Delta}=0$ in this case). Also,
given that $p_{+}=p_{-}$ for particle-antiparticle pairs, only part of
Biermann-battery effect survives (the relativistic -- carried by the fourth
term in the right-hand side of the above). This and the Joule-heating term
are the only purely relativistic corrections.

\section{Ohm's law for cold proton-electron plasmas}

\label{sOLCP-EPs} 

\subsection{Non-relativistic limit}

\label{ssN-RL} 
Cold plasmas have components with non-relativistic relative velocities.
Thus, at the low velocity limit, one can ignore terms of quadratic (and
higher) order in $v_{(i)}^{a}$. As a result, $\gamma_{(i)}\simeq 1$, $\hat{n}%
_{(i)}\simeq n_{(i)}$, $\hat{\mu}_{(i)}\simeq\mu_{(i)}$ and $\hat{h}%
_{(i)}\simeq\mu_{(i)}+p_{(i)}\simeq\mu_{(i)}$. If, in addition, the plasma
is a mixture of protons and electrons, we have $Z_{\pm}=\pm1$. When the
condition of overall electrical neutrality is also imposed, we may set $%
n_{\pm}=n$ and $\rho_{\pm}=\pm en$. Then, assuming that $m_{+}$ and $m_{-}$
are the proton and the electron masses respectively (with $\mu_{\pm}=nm_{\pm}
$ and $m_{-}\ll m_{+}$), we may write 
\begin{equation}
\hat{M}= \hat{M}_{+}+ \hat{M}_{-}\simeq nm_{+} \hspace{15mm} \mathrm{and} 
\hspace{15mm} \hat{\Delta}= \hat{M}_{+}- \hat{M}_{-}\simeq nm_{+}\,.
\label{nrhMhDel1}
\end{equation}
Note, however, that 
\begin{equation}
\hat{M}^{2}- \hat{\Delta}^{2}= 4\hat{M}_{+}\hat{M}_{-}\simeq
4n^{2}m_{+}m_{-}\neq 0\,.  \label{nrhMhDel2}
\end{equation}

Applying the non-relativistic limit to (\ref{rOhm1}) immediately removes the
last four terms from the right-hand side of the latter (all quadratic in $%
v_{\pm}^{a}$). For cold species we may also ignore the $\dot{\hat{p}}$ --
terms, though the spatial variations of the pressure are not necessarily
negligible. When dealing with proton electron systems, the eighth ($\mathcal{%
Q}_{+}$) term in the right-hand side of Eq.~(\ref{rOhm1}) vanishes (this is
clearly shown in (\ref{rOhm2}) -- recall that $Z_{\pm}=\pm1$). We may also
ignore energy-density changes due to collisions between the cold components,
and therefore remove the ninth term from the right-hand side of (\ref{rOhm1}%
). Finally, when applied to our two-component fluid, the collisional ($%
\mathcal{G}_{-}^{\langle a\rangle}$) term in (\ref{rOhm1}), which triggers
changes in the momentum of the species, takes the more familiar form 
\begin{equation}
-{\frac{e\hat{n}_{E}(\hat{M}_{+}+\hat{M}_{-})} {\hat{M}_{+}\hat{M}_{-}}}\, 
\mathcal{G}_{-}^{\langle a\rangle}= -\overline{\nu }\mathcal{J}_{a}= -\eta{%
\frac{e^{2}n}{m_{-}}}\mathcal{J}_{a}\,,  \label{6th}
\end{equation}
where $\overline{\nu}$ is the average collision frequency and $\eta=%
\overline{\nu}m_{-}/e^{2}n$ is the (scalar) electrical
resistivity of the two-component medium (e.g.~see~\cite{KT}). Note that,
although here we have adopted the common approximation of a scalar
electrical resistivity, our analysis also applies to general fluids with
anisotropic (tensor) resistivity. This can be done by using kinetic theory
to specify the interaction terms. On these grounds and by using the
auxiliary relations (\ref{nrhMhDel1}), (\ref{nrhMhDel2}), expression (\ref%
{rOhm1}) reduces to 
\begin{eqnarray}
\dot{\mathcal{J}}^{\langle a\rangle}&=& -\frac{4}{3}\,\Theta\mathcal{J}^{a}-
\left(\sigma^{a}{}_{b}+\omega^{a}{}_{b}\right)\mathcal{J}^{b}+ {\frac{e^{2}n%
}{m_{-}}}\,E_{a}+ {\frac{e^{2}n}{m_{-}}}\,
\varepsilon^{a}{}_{bc}v_{E}^{b}B^{c}- {\frac{e}{m_{-}}}\,
\varepsilon^{a}{}_{bc}\mathcal{J}^{b}B^{c}  \notag \\
&&+{\frac{e}{m_{-}}}\,\mathrm{D}^{a}p_{-}- \eta{\frac{e^{2}n}{%
m_{-}}}\,\mathcal{J}^{a}\,,  \label{nrOhm1}
\end{eqnarray}
in agreement with the expressions found in the standard literature
(e.g.~compare to Eq.~(3.5.9) in~\cite{KT}). Alternatively, we may recast the
above into the more familiar form 
\begin{equation}
E^{a}+ \varepsilon^{a}{}_{bc}v_{E}^{b}B^{c}- \eta\mathcal{J}^{a}= {\frac{%
m_{-}}{e^{2}n}} \left[\dot{\mathcal{J}}^{\langle a\rangle}+\frac{%
4}{3}\,\Theta\mathcal{J}^{a} +\left(\sigma^{a}{}_{b}+\omega^{a}{}_{b}\right) 
\mathcal{J}^{b}\right]+ {\frac{1}{en}}\, \varepsilon^{a}{}_{bc}\mathcal{J}%
^{b}B^{c}+ {\frac{1}{en}}\,\mathrm{D}^{a}p_{-}\,,  \label{nrOhm2}
\end{equation}
which immediately shows the terms responsible for the Hall and the
Biermann-battery effects -- see the last two terms in the right-hand side.
Either of these two expressions provides the 1+3 covariant form of the
generalised Ohm's law for a cold proton-electron plasma.

\subsection{Magnetohydrodynamic limits}

\label{ssMHDL} 
Adopting the commonly used magnetohydrodynamic (MHD) approximations, namely
ignoring all the right-hand side terms in Eq.~(\ref{nrOhm2}), leads to the
usual form of Ohm's law for an electrically resistive medium 
\begin{equation}
E^a+ \varepsilon^a{}_{bc}v_E^bB^c= \eta\mathcal{J}^a\,.  \label{Ohm5}
\end{equation}
We note here that, instead of dropping the 3-current terms seen inside the
square brackets of (\ref{nrOhm2}), one could in principle incorporate them
within the resistive term in the left-hand side of that expression. Then, we
will be referring to a ``generalised resistivity'' that accounts for changes
in the electrical properties of the medium due to relative motion
(i.e.~inertial) effects. It is also clear that in the idealised case of a
perfectly conducting medium, namely for $\eta\rightarrow0$, Eq.~(\ref{Ohm5})
reduces to the ideal-MHD form of Ohm's law. The latter is given by the
simple and well known formula 
\begin{equation}
E^a= -\varepsilon^a{}_{bc}v_E^bB^c\,.  \label{Ohm6}
\end{equation}

\section{Summary}

\label{sS} 
Hot plasmas are of major importance in a variety of physical phenomena,
ranging from laboratory physics to astrophysics and cosmology. A key factor
in determining the behaviour of plasmas is their electrical properties and
these are theoretically monitored by means of Ohm's law. The latter appears
in a number of different versions, which depend on the specifics of the
problem in hand. Here, we are providing fully relativistic and fully
nonlinear expressions for the generalised Ohm's law for plasmas, by deriving
the 1+3 covariant propagation equation of the 3-current density associated
with a multi-component fluid. Adopting a suitable definition for the
irreducible variables of the matter fields (see Eqs.~(\ref{Ie})-(\ref{Ih})
and (\ref{Ij}a)), we were able to address hot plasmas with a fluid-based
approach and without the need of kinetic theory. The use of the covariant
methods has also facilitated a mathematically compact and physically
transparent presentation of the subject. As a result, our expressions allow
for a direct comparison with the familiar Newtonian versions of Ohm's law,
while identifying the relativistic corrections to them. We show, for
example, that the relativistic analogue of the Biermann-battery effect has
an additional contribution from the temporal pressure variations. Our main
result is given in Eq.~(\ref{Ohm1}) and applies to any multi-component
fluid, relativistic or not, which means that it can be adapted to address a
great variety of physical problems. With the general form of Ohm's in hand,
our next step was to introduce a particular reference frame. Identifying our
fundamental observers with the Eckart frame, allowed us to follow on the
steps of the non-relativistic studies and therefore considerably simplify
the mathematics. Then, by confining to two-fluid systems and assuming
overall charge neutrality, we expressed Ohm's law in terms of the properties
of the bulk. Finally, we closed our discussion by considered a number of
applications. These included hot plasmas of two oppositely charged fluids,
hot electron-positron mixtures, cold electron-proton systems and also the
resistive and the ideal-MHD limits of our results. In each case, we have
discussed the physics of the situation, identified the familiar effects,
like the Biermann-battery and the Hall effects, and pointed to the
relativistic corrections were applicable.

The multi-fluid description adopted in the present paper is essential in
almost every study of small-scale astrophysical plasmas. The same approach
is also necessary when looking into the nonlinear regime of galaxy
formation, when the proto-structure has decoupled from the background
expansion and collapses. Then, one can use our equations to investigate the
evolution of proto-galactic magnetic fields, in particular their
amplification and dissipation, within and also outside the MHD limit.
Moreover, the use of the irreducible variables, assigns an unambiguous
physical interpretation to every variable in our equations and helps to
isolate the physical effects under consideration. For example, vorticity
terms are related to turbulence and dynamo-like mechanisms, while those
involving the shear describe shape distortions and can play an important
role during galaxy formation. In addition, going beyond the cold-plasma
limit, makes our equations suitable for studies of relativistic plasmas,
like those in hot interstellar clouds and in accretion discs around compact
stars. Applications of this sort will be the subject of future work.

\vskip0.5cm\textsl{Acknowledgements:} We thank Esteban Calzetta, Roy
Maartens and Loukas Vlahos for useful discussions and comments. A.K. also
acknowledges support from Projects PROPP-UESC 00220.1300.489 and
00220.1300.609.

\appendix

\section{The Landau frame}

\label{sLF} 
As stated in \S ~\ref{ssEFvsLF}, an alternative frame choice is that of the
Landau (or energy) frame. Both the particle and the energy frames are
physically equivalent and choosing one against the other depends on the
particulars of the problem in hand. The 4-velocities associated with the two
frames are given by (\ref{Eua}) and (\ref{Lua}) respectively, but rewriting
Ohm's law relative to the Landau frame goes beyond the scope of this paper.
We note, however, that the energy frame seems a less efficient choice,
because both the interpretation of the resulting equations and their
comparison to the non-relativistic expressions are less straightforward.
Nevertheless, by projecting $u_{L}^{a}$ along $u_{E}^{a}$, we can show that
the same non-relativistic limit will be attained. Indeed, definitions (\ref%
{Eua}), (\ref{Lua}) combine to 
\begin{equation}
u_{E}^{a}u_{a}^{L}= \frac{1}{\sum\limits_{i,j}n_{(i)} \sqrt{\hat{\mu}%
_{(j)}^{2}-\hat{q}_{(j)}^{2}}}\, \sum\limits_{i,j}\left[-\hat{n}_{(i)}\hat{%
\mu}_{(j)}+ \mathcal{N}_{(i)}^{a}\hat{q}_{a}^{(j)}\right]\,.  \label{VIt}
\end{equation}%
When the velocities of the individual species are small, we can neglect
terms proportional to the heat flux. This means that for cold plasmas the
above reduces to $u_{E}^{a}u_{a}^{L}\simeq \sum\limits_{i,j} \left[-\hat{n}%
_{(i)}\hat{\mu}_{(j)}\right]/ \sum\limits_{i,j}\left[\hat{n}_{(i)}\hat{\mu}%
_{(j)}\right]=-1$. Then, following (\ref{betai}), the Eckart and the Landau
frames (and their associated expressions of Ohm's law) effectively coincide.

\end{document}